\documentclass[%
preprintnumbers,twocolumn,
  superscriptaddress,
nofootinbib,
 amsmath,amssymb,
 aps,
 prl,
floatfix,
]{revtex4-2}

\usepackage{graphicx}
\usepackage{dcolumn}
\usepackage{bm}
\usepackage{xcolor}
\usepackage[version=4]{mhchem}

\begin{document}

\title{ Superpotentials, flat bands and the role of Quantum Geometry for the superfluid stiffness }

\author{T. Bauch}
\affiliation{Quantum Device Physics Laboratory, Department of Microtechnology and Nanoscience, Chalmers University of Technology, SE-41296 G\"oteborg, Sweden}
\author{F. Lombardi}
\affiliation{Quantum Device Physics Laboratory, Department of Microtechnology and Nanoscience, Chalmers University of Technology, SE-41296 G\"oteborg, Sweden}
\author{G. Seibold}
\affiliation{Institut f\"ur Physik, BTU Cottbus-Senftenberg, D-03013 Cottbus, Germany}

\begin{abstract}
  \textbf{ Enhancing superconductivity through material design is a central goal in quantum materials research. Moiré engineering, where twisting stacked layers creates long-wavelength modulations and flat bands, has shown how electronic correlations can be amplified  and eventually used to raise the superconducting critical temperature T$_c$. Yet this approach is largely confined to van der Waals materials and offers limited tunability. Here we explore a moiré-inspired alternative: imposing artificial superpotentials on otherwise homogeneous systems to engineer flat electronic minibands. Whether such superlattice potentials can truly enhance superconductivity and sustain a finite superfluid stiffness remains, however, an open question. Our calculations  show that a periodic superpotential  imposed to a 2D system can indeed enhance superconductivity by reconstructing the electronic bands and creating regions of large density of states, leading to a substantial increase of T$_c$.  In contrast to conventional  flat band systems, where the superfluid stiffness arises solely from quantum geometry through the quantum metric, a modulated system inherits kinetic energy from the filled minibands below the Fermi level. This inherited component coexists with a positive quantum geometric contribution, yielding a finite and robust stiffness even when the upper band becomes nearly flat. The resulting superconducting state remains coherent and resilient against weak to moderate disorder.
Our findings demonstrate that engineered superpotentials offer a tunable route to enhance superconductivity beyond twist based moiré systems, unifying flat band amplification of pairing with preserved phase stiffness. They further highlight the central role of quantum geometry in shaping collective electronic phenomena and point to superlattice design as a promising platform for next-generation superconductors.} 
\end{abstract}

\maketitle

When the magic-angle condition in twisted graphene was predicted \cite{macdonald11}, it opened an unexpected path in condensed matter physics. A subtle geometric twist between two layers produced electronic flat bands and, remarkably, superconductivity \cite{castro07,cao18,cao182,lu19,sharpe19,kerelsky19,yankowitz19,balents20} in a system where electrons were thought to be almost immobile. This discovery triggered an intense theoretical effort to understand how superconductivity could arise when the effective mass is infinite and the conventional contribution to the superfluid weight—linked to carrier motion—vanishes.
The key to this puzzle emerged  from the  quantum geometry concept. It was realized that in flat bands the superfluid stiffness can receive a finite contribution from the quantum metric \cite{toermae15,toermae16,toermae17,nagaosa21,bernevig22,toermae23,yang25}, 
a geometric property of the Bloch wavefunctions, rather than from the band dispersion itself. This insight transformed our understanding of superconductivity, revealing that the wavefunction’s geometry, not just its energy, determines the ability of the  condensate to support supercurrents.

Today, the concept of quantum geometry extends far beyond moiré materials and plays a fundamental role in a variety of condensed-matter phenomena \cite{gao25,verma25}.  For example it has emerged as a key ingredient in systems with non-trivial topology \cite{toermae15,peotta25,shavit24}, in multiband superconductors \cite{barlas24,hu25}, and in oxide interfaces such as LaAlO$_3/$SrTiO$_3$ \cite{sala25} where a finite quantum metric is associated with spin–momentum locking  which provides a tangible example of how real materials can harness geometric effects. Moreover, the quantum geometry of the electronic states can also influence the strength of the interactions like the electron–phonon coupling: bands with a larger quantum metric exhibit enhanced coupling to lattice vibrations \cite{yu24} which may in turn  boost superconductivity.  Indeed, flat bands themselves offer a tantalizing prospect: theory predicts that the superconducting transition temperature T$_c$ can scale linearly with the pairing strength $U$, rather than being exponentially suppressed as in the BCS limit \cite{shaginyan1,shaginyan2,volovik11,volovik18}. This opens a route to dramatically enhanced superconductivity, potentially even at ambient conditions, if the right quantum geometry is engineered. But must one  always rely on twisting between 2D layers to achieve flat bands? While twisting is confined to van der Waals materials,  the underlying idea of a moiré-like superpotential could, in principle, be engineered in a much broader class of superconductors.

\begin{figure*}[htb]
\includegraphics[width=10cm,clip=true]{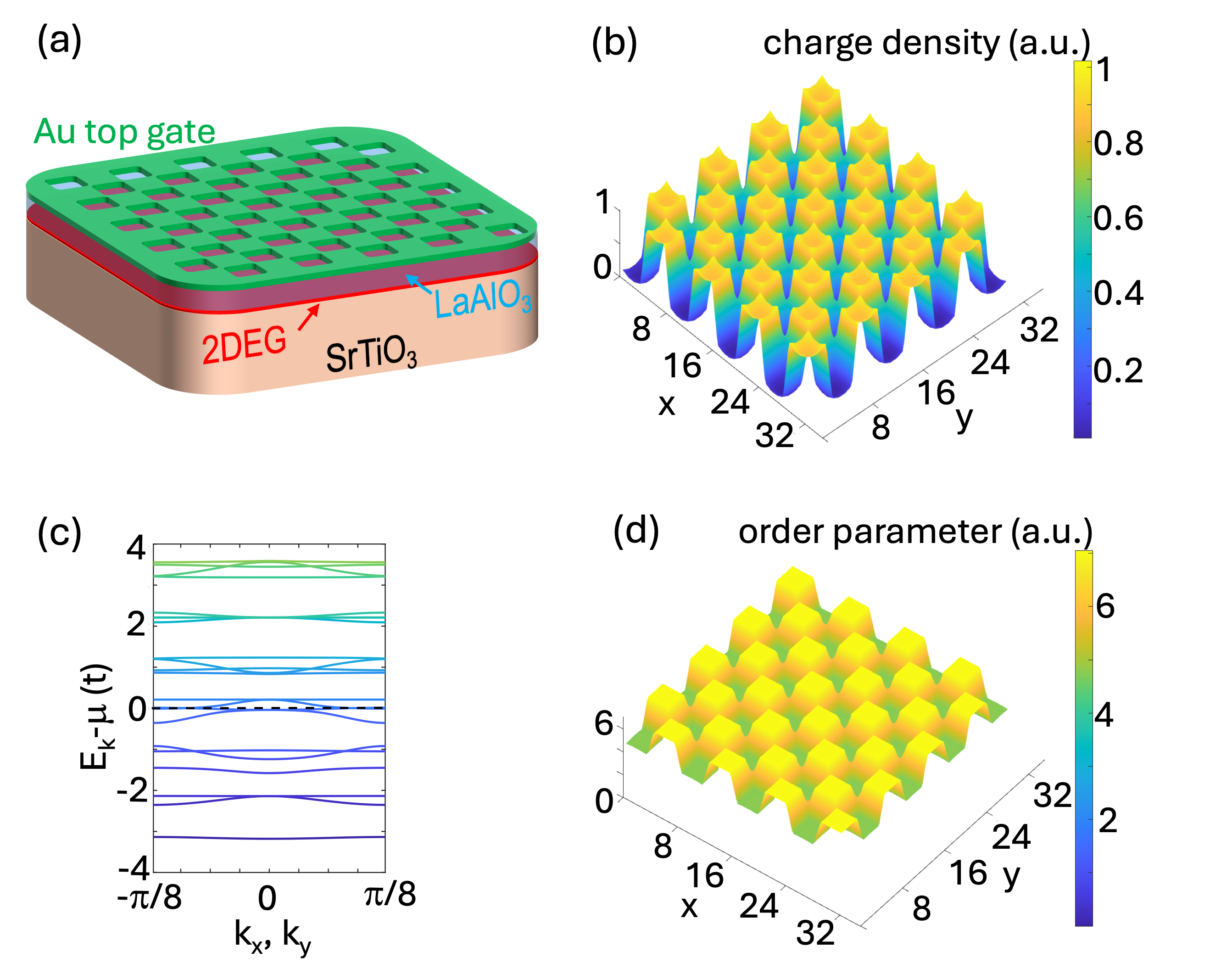}
\caption{{\bf Realization and electronic properties of a device with periodidic gating.} (a) Schematic of a 2D superconducting electron gas modulated by a periodic gate potential. A representative platform for realizing such a system is the LaAlO$_3$/SrTiO$_3$ interface. (b) Color surface plot of the modulated charge density for a checkerboard lattice of $4\times 4$ plaquettes. (c) Band structure for a diagonal cut across the reduced Brillouin zone. (d) Color surface plot of the modulated order parameter. Parameters: $U/t=1.5$, $V_0/t=4 (c)$, $V_0/t=2$ (b,d), charge concentration $p=0.5$. }
\label{fig1}     
\end{figure*}

When a conventional single band superconductor is subject to a periodic superpotential, the folded reduced Brillouin zone hosts minibands that can flatten at sufficiently high potential strength. In this regime, a key question arises: can the essential ingredients of a moiré-based superconductivity—enhanced density of states, elevated $T_c$, and finite superfluid stiffness—also emerge without geometric twisting?

Traditionally, a superpotential has been viewed as a weak perturbation, incapable of dramatically reshaping the electronic structure or inducing strong-coupling effects. Even when flat bands form at large  superpotential amplitudes at the Fermi energy E$_F$, they are typically not isolated, unlike in twisted graphene at the magic angle, but the remaining folded (and eventually dispersive) bands below E$_F$ still contribute to the response of the system. As a result, it has remained unclear whether band folding alone, i.e. in the absence  of the geometric effects of twisting, can produce flat-band physics and superconductivity enhancement.

Here, we demonstrate that contrary to this conventional view, a sufficiently strong superpotential, can indeed generate a flat-band system with a markedly enhanced superconducting transition temperature, $T_c$. Probing the superfluid stiffness reveals that, while quantum geometry  crucially contributes, its dominant component arises from the residual band dispersion of the occupied states below the Fermi level. Thus, even when the band at the Fermi energy becomes flat, the total kinetic energy remains finite, inherited from lower minibands.  This mechanism—absent in twisted moiré systems—establishes a new route to robust superconductivity in superpotential-driven flat-band materials and can readily be tested in two-dimensional superconductors  as illustrated in Fig. 1a.

\section{Electronic multiband structure and critical temperature}
\begin{figure}[h]
\includegraphics[width=6cm,clip=true]{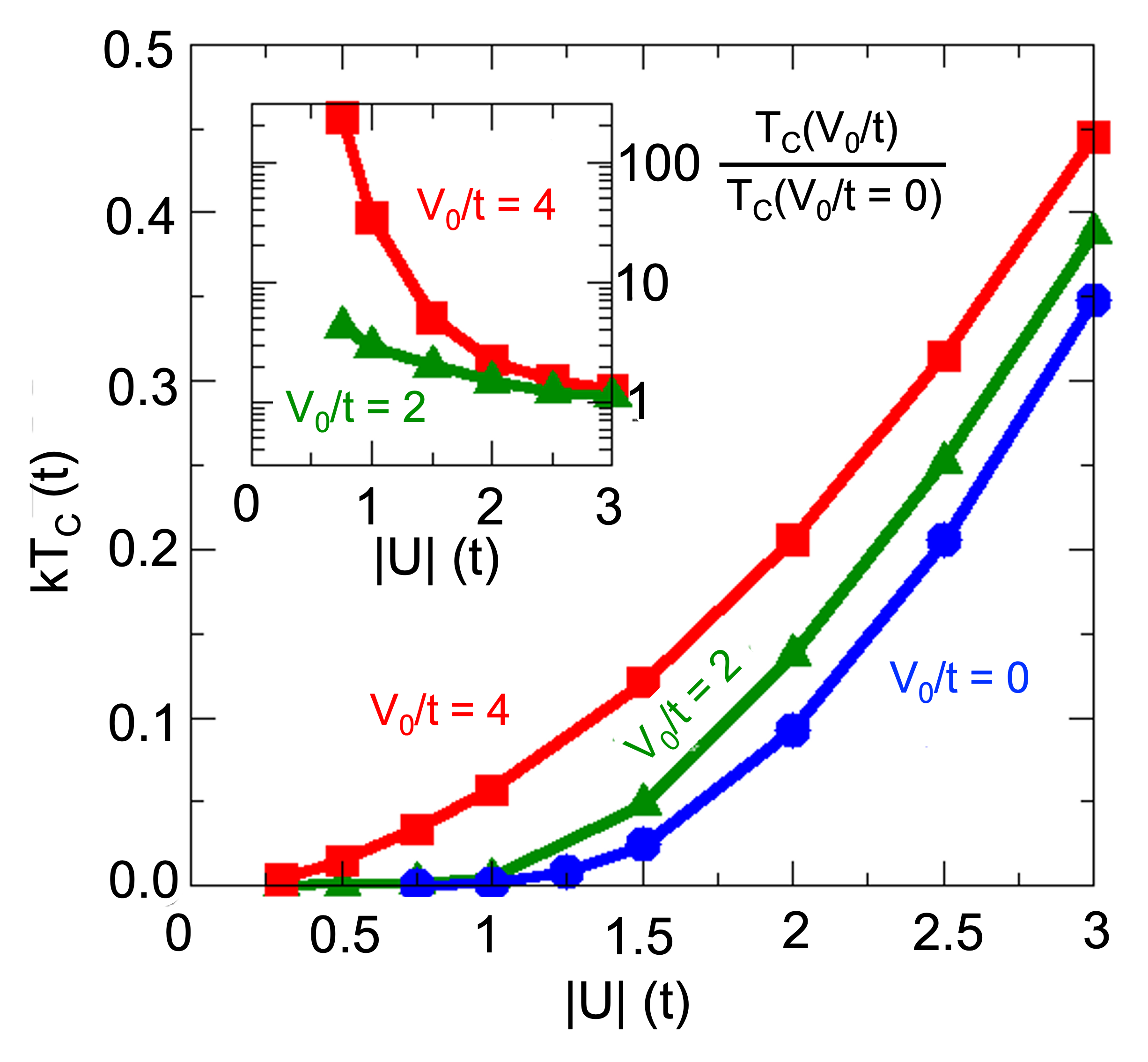}
\caption{{\bf Critical temperature in the presence of a superpotential.} Comparison of $T_c$ between homogeneous system ($V_0/t=0$, blue) and in the presence of a superpotential with $V_0/t=2$ (green) and $V_0/t=4$ (red). Inset: Ratio between $T_c$ with a finite superpotential and $T_c$ for the homogeneous system. The solid red line in (e) is a fit to $T_c/T_c^{hom}=\alpha |U| e^{\beta t/|U|}$ with $\alpha=0.0436$ and $\beta=6.663$ as described in the text. Charge concentration: $p=0.5$.}
\label{fig2}     
\end{figure}

Our calculations are based on the following tight-binding hamiltonian 
\begin{eqnarray}
  H&=&-t\sum_{\langle ij\rangle,\sigma} c_{i,\sigma}^\dagger c_{j,\sigma}
  +\sum_{i\sigma}\left(V_i-\mu\right) n_{i\sigma} \nonumber \\
  &-&|U|\sum_n \left\lbrack n_{n,\uparrow}-\langle n_{n,\uparrow}\rangle\right\rbrack
  \left\lbrack n_{n,\downarrow}-\langle n_{n,\downarrow}\rangle\right\rbrack\label{eq:hi}\,,
\end{eqnarray}
which contains a local attraction $-|U|$ between charge
fluctuations and we restrict to hopping between
nearest neighbor sites ${\bf R}_i$, ${\bf R}_j$
indicated by $\langle ij\rangle$.
For the local potential $V_i$ we consider a checkerboard geometry with
alternating $4\times 4$ plaquettes within which $V_i$ varies between
the values $V_i=0$ and $V_i=-V_0$. Eq. (\ref{eq:hi}) is solved via a standard BCS mean-field decoupling, see Supplementary Section I for details. 
Charge density and order parameter are
enhanced on the plaquettes with $V_i=-V_0$ (Fig. \ref{fig1}b and d) and the resulting minibands display a flat character for sizeable $V_0$, Fig. \ref{fig1}c. The corresponding increased density of states induces a substantial enhancement of $T_c$, in particular at weak coupling, Fig. \ref{fig2}. In fact, for flat bands $T_c \sim |U|$ \cite{shaginyan1,shaginyan2,volovik11,volovik18} while for dispersive bands the standard BCS relation 
$T_c \sim exp(-1/(N(0){|U|}))$ applies, where $N(0)$ denotes the density of states at the Fermi level.

\section{Superfluid stiffness}

The superfluid stiffness $D_s$ along a given direction $\alpha$ is calculated from the current correlation function $\chi^{\alpha\alpha}_{jj}$ \cite{scala93}
\begin{equation}
D_s=-\langle T_\alpha\rangle - Re \chi^{\alpha\alpha}_{jj} \label{eq:ds}
\end{equation}
defined in the momentum limit $q_{\parallel}=0, q_{\perp}\to 0, \omega=0$
corresponding to the Meissner effect.
It is composed of a diamagnetic part, in the present
model equivalent
to the negative kinetic energy $-\langle T_\alpha\rangle$ (that for parabolic dispersions equals the conventional $n_s/m$), and a paramagnetic term, which arises in inhomogeneous systems.
Within our model this term contributes due to the presence of the superpotential scattering. Here we need to point out that contrary to conventional flat band systems we start from a model which has a finite kinetic energy. The multiband structure arises from the folding of the original band and splitting of the minibands due to the superpotential. Thus, even if the kinetic energy is reduced 
by the superpotential, it generally will remain finite, even if the miniband which crosses $E_F$ becomes completely flat.

We also explore a  parameter regime, where the interaction
$|U|$ becomes larger than the hopping $t$. In this regime one cannot neglect the contribution of phase and charge fluctuations to the superfluid stiffness. These fluctuations take into account the phase relaxation of the SC order parameter in response to the applied vector potential and renormalize the stiffness for the inhomogeneous system \cite{seibold12}. 
\begin{figure}[h]
\includegraphics[width=6cm,clip=true]{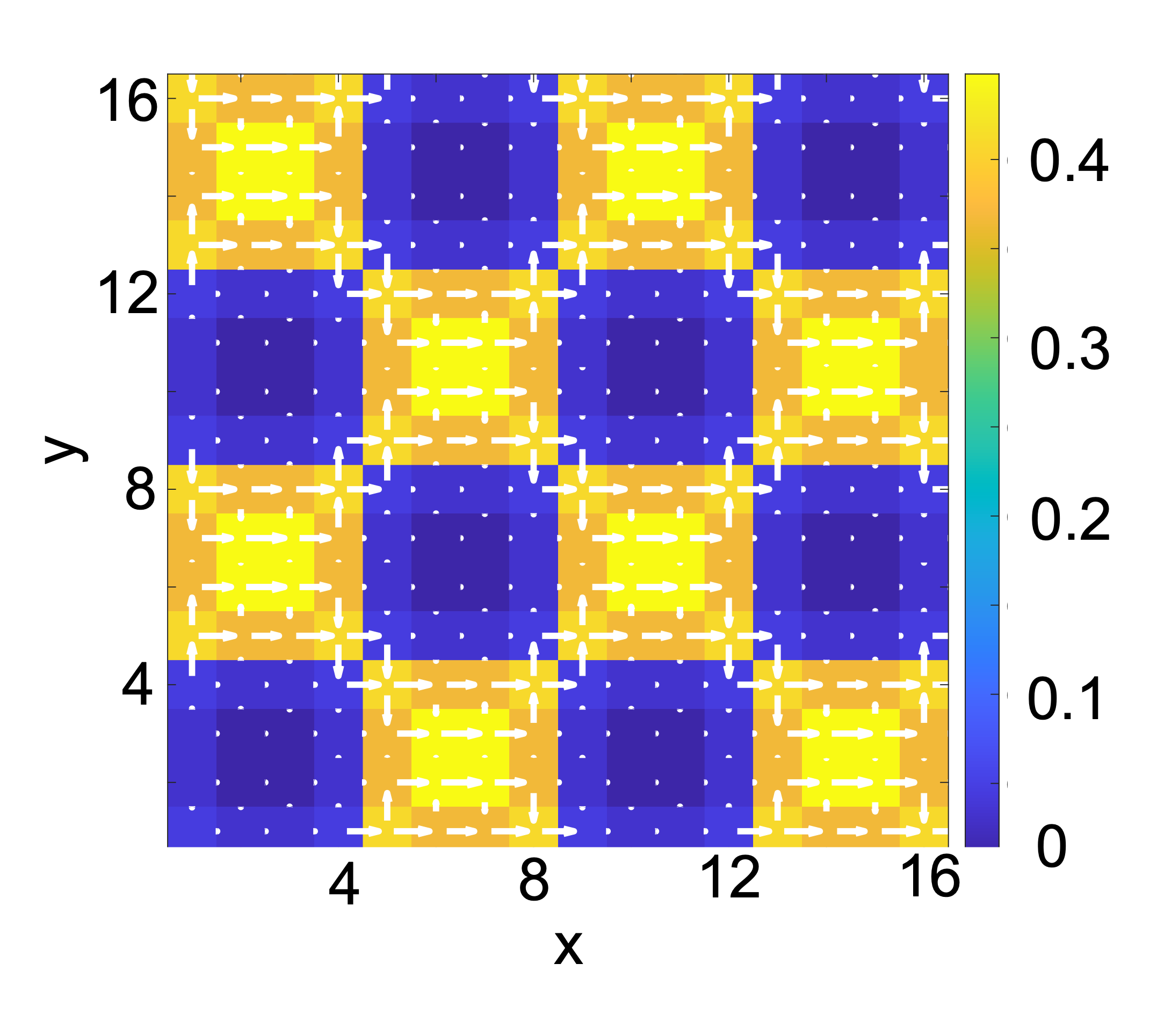}
\caption{{\bf Current flow through the periodically modulated superconductor.} Illustration of the gauge invariant SC current distribution (white arrows) for a vector potential applied along the x-direction. The local gaps are shown as colored background.  $U/t=2$, $V_0/t=4$, Charge concentration $n_{el}=0.5$.}
\label{fig3} 
\end{figure}
Formally, this implies that the BCS current correlations have to be dressed by collective charge and order parameter fluctuations, cf. Supplementary Section IV.
The total stiffness $D_s=D_s^{BCS}+D_s^{RPA}$ is composed of a BCS and a fluctuation contribution
which we compute in the random phase approximation (RPA) \cite{seibold12}. This also ensures gauge invariance so that the SC currents are conserved at
each node of the lattice, as shown in  Fig. \ref{fig3}.
One observes, that currents are dominant on the plaquettes with large SC order parameter and connect these plaquettes diagonally via currents along the y-direction. 
As shown in Supplementary Figure 2, within the BCS approximation instead one finds violation of charge conservation.

\begin{figure}[ttt!]
\includegraphics[width=8cm,clip=true]{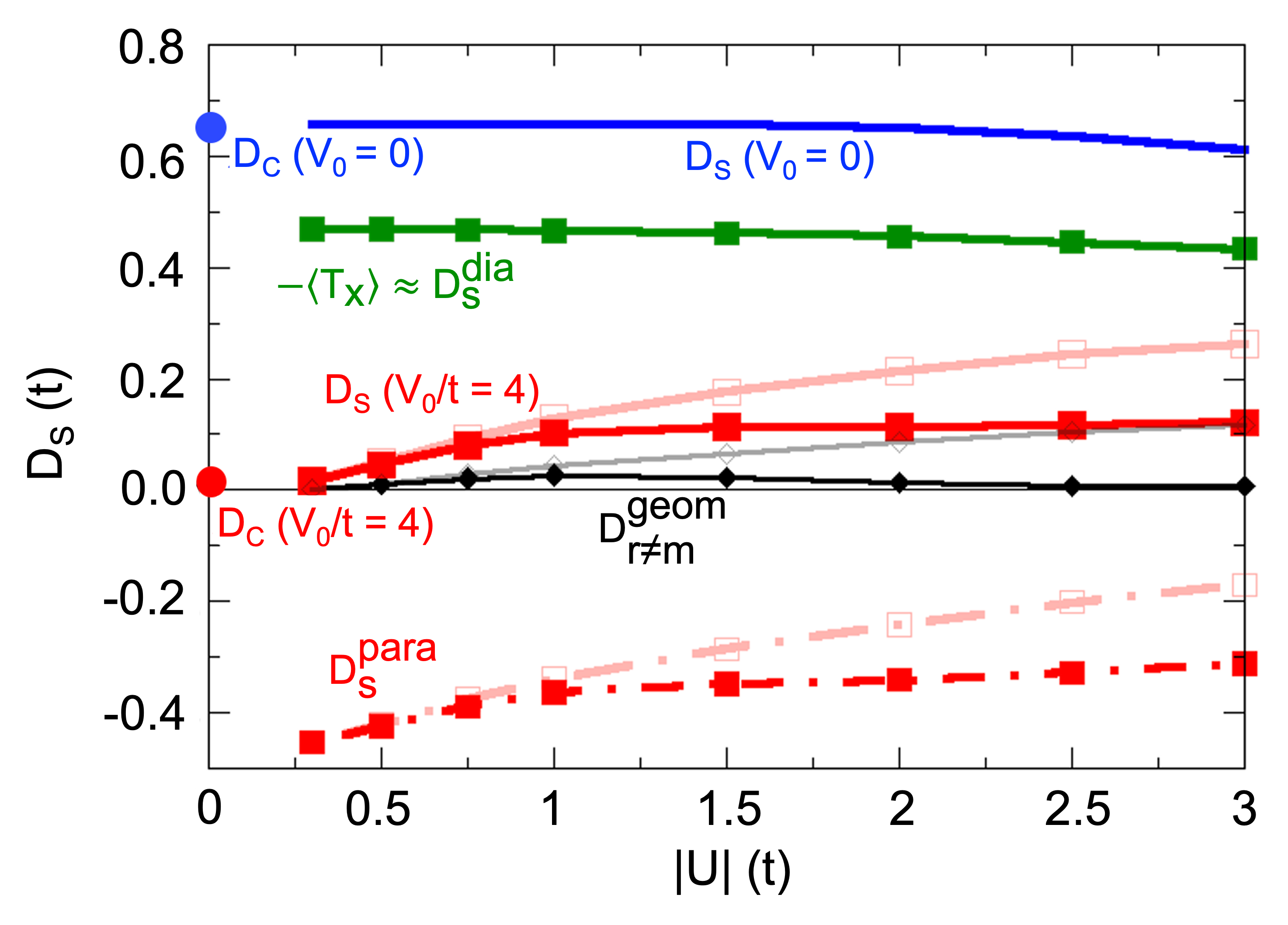}
\caption{{\bf Impact of superpotential and gauge invariance on the superfluid stiffness.} Interaction dependence of the stiffness $D_{s,tot}$for superpotential values $V_0/t=0$ (blue) and $V_0/t=4$ (red, squares) separated into the diamganetic (mostly intraband)
($D_{s}^{dia}$) and paramagnetic (mostly interband) ($D_{s}^{para}$) contribution.  Also shown is the geometric contribution for $r\ne m$ (black) which positively contributes to the stiffness. Lines in the foreground (full symbols) and in the background (open symbols) refer to the RPA (BCS) result, respectively. For $|U|/t=0$ the charge stiffness $D_ c$ is shown by full circles while $D_s$ vanishes in this limit. charge concentration: $p=0.5$.}
\label{fig55}
\end{figure}

Fig. \ref{fig55} shows the interaction dependence of $D_s$ for $V_0/t=4$  in comparison
with the homogeneous case ($V_0/t=0$) (blue line) within the
BCS (  light red line)   and BCS+RPA approach (red full line), respectively.
In the limit $|U|/t\to 0$, $D_s$ approaches the Drude weight $D_c$, defined similarly to Eq. (\ref{eq:ds}) but in the limit
${\bf q}=0, \omega\to 0$ \cite{scala93}.
At $U/t=0$ the stiffness $D_s=0$ both for homogeneous and non-homogeneous systems.

 For the homogeneous system  one observes a slight reduction of the stiffness with increasing attractive interaction. Since in this case $D_s$ is only due to the
 diamagnetic part, the behavior of the stiffness is solely determined by the increase of the kinetic energy (i.e. becoming less negative)
with increasing order parameter which is the usual behavior for a BCS superconductor~\cite{bcs}.
Upon increasing the superpotential $V_0$ the overall stiffness gets suppressed with respect to the homogeneous case due to the increasing paramagnetic contribution.
In particular, we find that for $V_0/t=4$ and $|U|/t<0.5$ the
stiffness almost vanishes ($D_s\approx 0$) which also holds for the charge stiffness ($D_c\approx 0$).  For smaller values of
$V_0/t$ one still finds a finite stiffness in the limit $|U|/t\to 0$, see Supplementary Figure 3, while the enhancement of $T_c$ is much less pronounced, see Fig. \ref{fig2} (green curve). In case $V_0/t=4$ it is therefore remarkable
that upon increasing the interaction the stiffness partially recovers and
for $|U|/t=3$ approaches a value of $D_s/t\approx 0.1$ (red full line) which is almost $17\%$ of the
value for the homogeneous system (blue line)  so that $\pi D_s$  is of the order of $kT_c$, cf. Fig. \ref{fig2}. Note also that the inclusion of phase relaxation via the BCS+RPA approach induces a significant reduction of the paramagnetic contribution to the stiffness from its BCS value (light red curve)  for $|U|/t\gtrsim$ 1.
While the interaction dependence of $D_s$ for $V_0/t=4$ mimics the behavior one would expect
from the geometric contribution in a flat band SC \cite{toermae16,nagaosa21}, 
here the recovery of $D_s$ is based on a different mechanism that, however, also includes a positive geometrical contribution.
First, let us note that the kinetic term $-\langle T_\alpha\rangle$  (green line) stays finite due to the contribution of all occupied minibands as explained before. 
Second, the negative paramagnetic contribution (dashed-dotted line), which for $V_0/t=4$ compensates the diamagnetic one at small $|U|/t$, becomes less negative upon increasing the attractive interaction. 
This can be understood based on the transfer of low energy spectral weight of the optical conductivity to its $\omega=0$ pole with
increasing SC gap as it is found in diffusive systems, see Supplementary Section VII.
Nevertheless, we will show in the following that a certain geometric contribution is involved in this transfer of spectral weight.

This analysis is based on separation of the stiffness 
in a conventional (intraband)
and a geometric (interband)
contribution. The latter depends on the (real part of the) quantum metric tensor
$g^{mn}_{\alpha\beta}= \langle n|\partial k_\alpha |m\rangle \langle m|\partial k_\beta |n\rangle $
\cite{nagaosa21} which is associated with the
interband matrix elements of the current operator.
The geometric contribution therefore manifests as a superconductivity induced
spectral weight transfer in the optical conductivity of a multiband system, cf. Supplementary Section VI, and constitutes the dominating contribution to the stiffness in the conventional flat band limit.

We therefore rewrite the stiffness in the diagonal basis, cf. \cite{toermae17} which for the 
BCS contribution reads
\begin{eqnarray}
D^{BCS}_s&=&-\frac{1}{N}\sum_{\substack{k,r,t,n,m\\ p_{<},p_{>}}}\hspace*{-0.3cm}{}^{'}\left\lbrace
\tau^\alpha_{r,t}(k,p_{<},p_{>})\delta_{rn}\delta_{tm} \right.\nonumber \\
&+&\left. 2\frac{C^\alpha_{r,t}(k,p_{<},p_{>}) C^{\alpha,*}_{m,n}(k,p_{<},p_{>})}{\Omega_{p_{>}}+\Omega_{p_{<}}}\right\rbrace \label{eq:dsbcs}
\end{eqnarray}
where the indices $r, t, n, m$ label the (normal state) miniband, and $p_{<(>)}$ refer to the
quantum states in the SC state with energies $\Omega_{p_{<(>)}}$.
  $\tau^x$ is the diamagnetic contribution to the stiffness, expressed within these basis states, and $C^\alpha$ refer to the matrix elements of the current
  operator explicitely given in Supplementary Section III.
  The RPA-contribution can be cast into a similar
  form
\begin{eqnarray}
  D^{RPA}_s&=&
\frac{1}{N}\sum_{\substack{k,r,t,n,m\\ p_{<},p_{>},p'_{<},p'_{>}}}\hspace*{-0.3cm}{}^{'}
C^\alpha_{r,t}(k,p_{<},p_{>}) \label{eq:dsrpa}\\ &\times&\Lambda^{RPA}(k,p_{<},p_{>},p'_{<},p'_{>})C^{\alpha,(*)}_{m,n}(k,p'_{<},p'_{>})\nonumber
\end{eqnarray}
and the RPA kernel $\Lambda^{RPA}$ is given in Supplementary Section IV.
The conventional (intraband) contribution $D^{intra}$ is obtained 
from Eqs. (\ref{eq:dsbcs},\ref{eq:dsrpa}) by restricting the
summation to $r=t$ and $n=m$ whereas the geometric
contribution $D^{inter}$ contains all the interband processes
with $r\ne t$ and $n\ne m$.

We further separate the stiffness into intra-, inter- and mixed band contributions:
\begin{eqnarray}
D_s&\equiv&\sum_{r,m}{\cal D}_{r,m}\equiv \sum_{r,t,n,m}{\cal D}_{r,t,n,m} \label{eq:defdrm}\\
{\cal D}_{r,m} &=& {\cal D}^{intra}_{r,r,m,m} + \sum_{t\ne r,n\ne m}{\cal D}^{inter}_{r,t,n,m} \label{eq:dsinter}\\
&+& \sum_{n\ne m}{\cal D}^{mixed(1)}_{r,r,n,m}
+\sum_{t\ne r}{\cal D}^{mixed(2)}_{r,t,m,m} \nonumber \,.
\end{eqnarray}
The latter correspond to processes where one of the two current matrix
elements $C^x_{r,t}$ in Eqs. (\ref{eq:dsbcs},\ref{eq:dsrpa}) describes intra- and the other one interband scattering. It turns out that for present parameters the mixed contributions are smaller by a factor $\sim 10^{-2}$ than ${\cal D}^{intra/inter}$ and therefore will be neglected in the following discussion.

\begin{figure}[htb]
\includegraphics[width=7cm,clip=true]{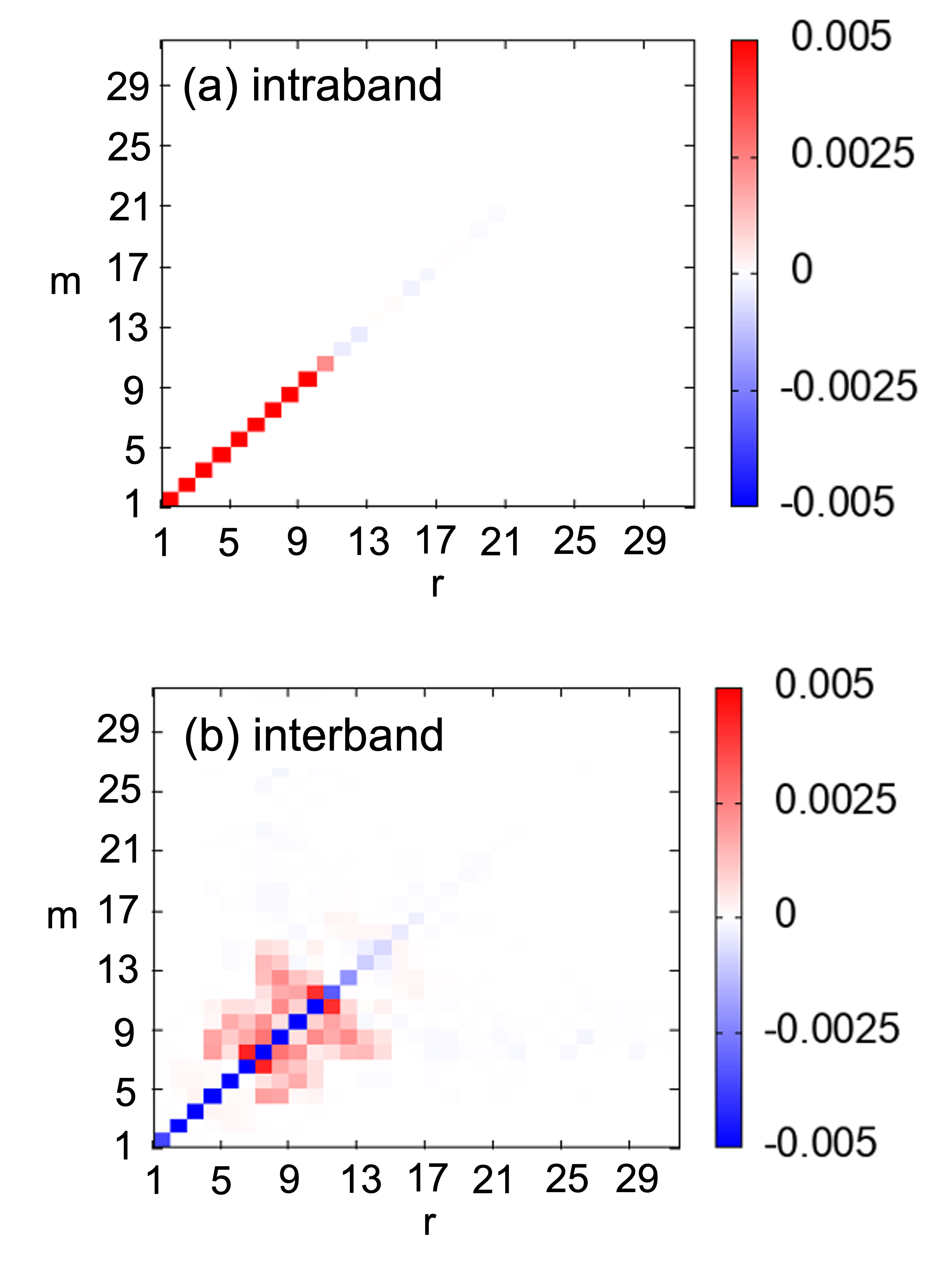}
\caption{{\bf Intra- and interband contributions to the
superfluid stiffness.} Color surface plot of ${\cal D}_{r,m}$, see
Eq. (\ref{eq:defdrm}), divided into intraband (a) and
interband (b) contributions. 
charge concentration $p=0.5$; $|U|/t=2$.}
\label{fig52}
\end{figure}

Fig. \ref{fig52}a shows ${\cal D}^{intra}_{r,r,m,m}$
where the diagonal elements $r=m$ are dominated by
the (negative) kinetic energy contribution of band $r$. For
the considered doping $p=0.5$ the lowest (occupied) $8$ bands 
contribute to the kinetic energy in the normal state.
Due to particle-hole mixing there is also a contribution from bands above the chemical potential in the presence of SC. The sum over all contributions in Fig. \ref{fig52}a is very close to the diamagnetic term, i.e. the total kinetic energy.
On the other hand, the sum over the interband (i.e. geometric) contributions $D^{geom}_{r,m}=\sum_{t,n}{\cal D}^{inter}_{r\ne t, n\ne m}$ in
Fig. \ref{fig52}b is always negative and is close to the (negative) paramagnetic term. It can be seen that this is dominated by diagonal elements  ${\cal D}^{geom}_{m,m}$ which means that in 
Eq. (\ref{eq:dsinter}) the two current matrix elements describe scattering processes which involve a common band $r=m$. Interestingly, we also find a positive interband term
$D^{geom}_{r\ne m}$ (cf. black curve in Fig. \ref{fig55}) 
which in conventional flat band systems
would constitute the dominating contribution to the stiffness. It scales with the interaction and represents a significant fraction of $20\%-30\%$
of the total stiffness in weak coupling while its
contribution becomes suppressed due to phase relaxation  for $|U|/t \gg 1$.

\section{Consideration of disorder}
Within our 'mean-field' BCS approach the critical temperature $kT_c$ can be
associated with the pairing energy scale whereas $\pi D_s$ sets the scale for phase coherence. 
In the weak-coupling BCS limit one therefore
always finds $\pi D_s/kT_c \gg 1$ since phase coherence is already established when Cooper pairs
start to form. In the low-disorder regime where
the mean free path is much longer than the inverse Fermi wave vector, $k_F l \gg 1$, the stiffness slightly decreases due to the paramagnetic contribution while 
Anderson's theorem \cite{anderson59} for the pairing of time-reversed exact eigenstates leaves $T_c$ unchanged. 
The situation changes when $k_F l\approx 1$ 
and Cooper pairs can start to form above the temperature where phase coherence sets in,
cf. e.g. Ref. \cite{mondal13}.
In this regime the stiffness becomes comparable
with $kT_c$ eventually leading to a superconductor-insulator transition ($D_s=kT_c=0$) when disorder exceeds a critical value as e.g. observed in $TiN$, $InO_x$ and $NbN$ films \cite{sac10,sac11,mondal11}.

Charge carriers in flat band systems are very susceptible toward localization 
so that the stiffness, which depends inversely on the
mass, also sensitively depends on the amount of disorder when the strength of the superpotential
becomes sizeable.

In order to investigate the influence of disorder we add to the hamiltonian Eq. (\ref{eq:hi}) an impurity potential
$V=\sum_{n\sigma}I_n n_{n\sigma}$ where the $I_n$ are drawn from a flat distribution with $-I_0 \le I_n \le I_0$. 
In this way disorder is treated exactly, cf. Supplementary Section II, and we evaluate the resulting stiffness by
averaging the current correlations over $10-20$ disorder configurations.

\begin{figure}[htb]
\includegraphics[width=8cm,clip=true]{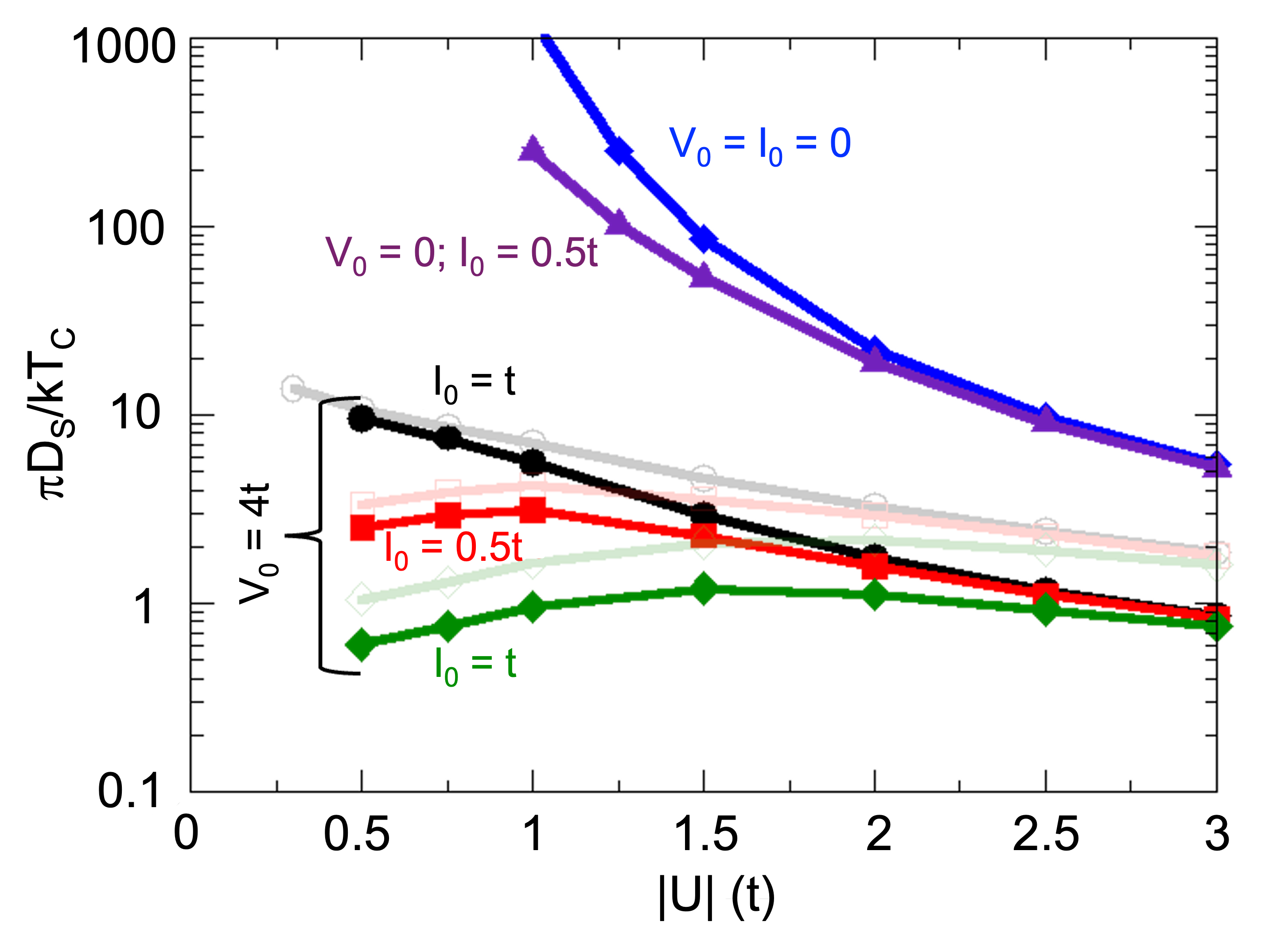}
\caption{{\bf Ratio between the superfluid stiffness and $T_c$.} Blue and violet curves are for the system without
  superpotential and impurity potentials $I_0/t=0.$ and $I_0/t=0.5$, respectively. In the presence of a superpotential ($V_0/t=4$, patch size $4\times 4$) the results are shown for impurity strength $I_0/t=0$ (black), $I_0/t=0.5$ (red), $I_0/t=1$ (green). Lines in the foreground (full symbols) and background (open symbols) refer to the RPA (BCS) result, respectively. In case of $V_0=I_0=0$ RPA and BCS result are identical, whereas for $V_0=0$, $I_0=0.5t$ the difference is within the size of the symbols. Charge density $n_{el}=0.5$.}
\label{fig6}                                          
\end{figure}  

 We consider two values for the impurity potential 
 $I_0/t=0.5$ and $I_0/t=1$ which, upon solving the temperature dependent gap equation, have only minor
 influence on $T_c$. The relaxation time is obtained by fitting $\sigma(\omega)$ to the Drude formula, see Supplementary Section V 
and $k_F$ is determined from the average of the Fermi momentum over the Fermi surface
of the homogeneous system. For $I_0/t=0.5$ this yields
a value of $k_F l\approx 40$ and a mean free path of $l\approx 23$ in units of the lattice spacing which is larger than the superlattice periodicity.
In case of $I_0/t=1$ one finds $k_F l \approx 10$ and $l\approx 6$ which is of the order of the
superlattice periodicity (we recall that for the $4\times 4$ plaquette structure the translation vectors are
$\vec{a}_1=(8,0)$ and $\vec{a}_2=(4,4)$ in units of the lattice constant).
 
Fig. \ref{fig6} compares $\pi D_s/(kT_c)$ for systems with and without
superpotential and impurities.
In the clean and homogeneous case ($V_0/t=I_0/t=0$, blue) the typical BCS behavior is recovered
where in the weak coupling limit the stiffness exceeds the pairing scale
$kT_c$ by several orders of magnitude and the ratio 
$\pi D_s/(kT_c)$ decreases
towards stronger coupling. In fact, since in the
homogeneous system the stiffness only weakly depends on $|U|$, the behavior can be understood from the
weak-coupling BCS result $T_c \sim e^{-1/(|U|N_0)}$
where $N_0$ denotes the density of states. This yields
$ln(D_s/T_c)\sim 1/|U|$ which is exactly the behavior
shown in Fig. \ref{fig6} for the homogeneous system.
For a moderate impurity strength ($V_0/t=0; I_0/t=0.5$, violet) the stiffness of the homogeneous system is reduced in particular at weak coupling while for larger interaction it approaches the value of the
clean case.

In the presence of a superpotential but without impurities ($V_0/t=4; I_0/t=0$, black) the ratio $\pi D_s/kT_c$ is substantially smaller than for the
homogeneous system, in particular at weak coupling. On the one hand, this is due to the smaller stiffness in this regime (cf. Fig. \ref{fig55}) but more relevant is the exponentially larger $T_c$, cf. inset to Fig. \ref{fig2}. While for the clean superpotential $\pi D_s/kT_c\approx 10$, one finds a further
reduction of this value at weak
coupling $|U|/t=0.5$ due to the addition of impurities.
We also observe that $\pi D_s/kT_c$ goes through
a maximum upon decreasing the interaction and
for impurity strength $I_0/t=1$ (green) it approaches 
$\pi D_s/kT_c\approx 1$ for $|U|/t=0.5$.
Thus, this limit is obtained when
the mean free path becomes of the order of the
superpotential periodicity.

\section{DISCUSSION}

The periodic superpotential can substantially enhance the critical temperature while preserving a sizeable stiffness that scales with the interaction strength. Unlike conventional flat-band systems, where filled bands do not contribute to the kinetic energy and a finite stiffness arises solely from the positive quantum geometric term, our periodically modulated system retains contributions from all minibands below the chemical potential. As a result, even when the top band is flat, a finite diamagnetic response persists. The interband term is overall negative but includes a positive quantum geometric component. Thus, superlattice engineering offers a promising route to enhance Tc, as the system effectively inherits the stiffness of the homogeneous case.
It is interesting that recent theoretical work 
succeeded to predict the topological properties of materials subject to a superlattice potential \cite{crepel25}. In particular,
the periodic patterning of a 2D electron gas \cite{naumis25} reveals a non-zero local Berry phase for antisymmetric triangular superpotentials. This geometry therefore could further sustain the superfluid stiffness because of topology. Consideration of such structures will be considered in future studies within the Usadel equations which allow for the treatment of superpotentials which are incommensurate with the host lattice.

It should be noted that the model here differs from previous work which studied the competition between charge-density wave and SC order parameters \cite{balseiro77,gabovich2002}.
Here instead, we are not considering competing instabilities since the superpotential is imposed on the system  via external periodic gating \cite{klochan24} or nanostructuring \cite{boosting} so that the Fermi level can in principle be tuned into one of the flat bands. This leads to
an enhancement of $T_c$ with respect to the homogeneous system which at weak coupling can reach
one or two orders of magnitude, cf. inset to Fig. \ref{fig2}.
It is to be expected that the influence of phase fluctuations will affect this amplification if the mean free path in the disordered system is of the same order of magnitude as the periodicity of the superlattice which corresponds to the limit $\pi D_s/kT_c \approx 1$. For strongly disordered NbN films \cite{mondal13} it has been shown that the transition temperature toward a coherent SC state is reduced by a factor $\sim 3$ when 
this ratio decreases from $\pi D_s/kT_c\approx 5$ to
$\pi D_s/kT_c\approx 1$ so that our results are still compatible with a significant enhancement of $T_c$.

Yet, there are still a few hurdles to overcome in order to realize such a system with increased transition temperature and robust stiffness. 
First, secondary instabilities such as magnetism may occur when the Fermi level is located in one of the flat bands.

The concomitant opening of additional gaps will then compete with the T$_c$ enhancement although the system may remain in a state with sizeable stiffness.
Second, as we have seen the system should be sufficiently clean so that the stiffness still dominates the energy scale of the transition temperature. 
In principle, this requirement should be achievable with 
 LaAlO$_3$/SrTiO$_3$ interfaces where tunable superconductivity has been observed below $0.3$ K \cite{biscaras12}. Here one can reach mobilities as high as $10.000$ $cm^2/Vs$, corresponding to a mean free path of about
$100$ nm \cite{singh}. This is well beyond the $40$ nm pitch limit of standard e-beam lithography which defines the
possible lower bound periodicity of the metal gate superpotential \cite{forsythe}.

  To conclude, this work demonstrates that applying a superpotential can provide a solid foundation for enhancing superconductivity, even in single-band systems. It thereby paves the way for engineering  diverse  superpotential architectures as a powerful strategy to design and control superconductivity.

\begin{acknowledgements}
The authors acknowledge support from the Swedish Research Council (VR), under the Projects 2018-04658 (F.L.) and 2020-05184 (T.B.), and by the Deutsche
Forschungsgemeinschaft, under SE 806/20-1 (G.S.).
\end{acknowledgements}

\end{document}